\documentstyle[aps,prl,epsf]{revtex}
\begin{document}
%\draft
\twocolumn[\hsize\textwidth\columnwidth\hsize\csname 
@twocolumnfalse\endcsname
\title{Wigner Glass, Spin-liquids, and the Metal-Insulator Transition}
\author{Sudip Chakravarty, Steven Kivelson, Chetan Nayak, and Klaus 
Voelker}
\address{Department of Physics and Astronomy\\University of 
California Los
Angeles\\Los Angeles, California 90095-1547}
\date{\today}
\maketitle
%\begin{abstract}
%\end{abstract}
%\pacs{}
\vskip 0.5 in 
]
\nopagebreak
{\bf Recent experiments on the two dimensional ($d=2$) electron gas 
in various semiconductor devices have  revealed an unexpected 
metal-insulator transition\cite{Krav1,Popovic,GaAs,SiGe}
and have challenged the previously held assumption
that there is no such transition in two dimensions. 
While the experiments are still at the
stage of rapid development, it is becoming  evident that they 
cannot be understood from the
conventional perspective of weak interactions.
In the present paper, we propose the following. (1) The low-density
insulating state is the Wigner Glass, a phase with quasi-long-range 
translational
order and competing ferromagnetic and antiferromagnetic
spin-exchange interactions.
(2) The transition is the  melting of this Wigner Glass,
disorder being the agent allowing the transition to be second order.
(3)  Within the Wigner Glass phase, there are at least two, distinct
magnetic ground-states, a ferromagnetic state at very low electron density 
and a spin-liquid state with a spin pseudo-gap at higher densities.
(4) The metallic side of the transition is 
a non-Fermi liquid.
These conclusions are encapsulated in Figure 1 which presents
the proposed phase diagram as a function
of disorder strength and density;  we also suggest 
experimental signatures of the various phases and transitions.}

Although there is not complete agreement between experiments, a 
number of striking features have
emerged. 
Most importantly, the temperature dependence of the resistivity 
changes
at a critical carrier concentration ($n_c$): for
lower carrier concentration ($n<n_c$), it is insulating, implying a 
diverging resistivity as the 
temperature, $T\rightarrow 0$; for higher carrier
concentration ($n>n_c$), it is metallic, the resistivity
decreasing with decreasing temperature. Therefore, $n_c$ appears to 
signify a zero temperature quantum phase 
transition between two
fundamentally distinct states of matter. It was  previously thought 
that such a metal-insulator transition is not 
possible in two dimensions\cite{Review}. Indeed,  previous 
experiments,
carried out over the past two decades on systems with higher electron 
density 
where the effects of electron-electron interactions are significantly 
weaker, strongly supported the
idea that a metallic state in two dimensions is not possible.

In some of the recent experiments, a striking reflection symmetry is 
observed\cite{Simonian}. It is striking because, 
{\em a priori\/}, there is no particular symmetry between the two 
distinct states of matter, the metal
and the insulator. Over a significant range of $n$ around $n_c$, one 
finds that 
$g^*(\delta_n,T)=1/g^*(-\delta_n,T)$,
where $g^*$ is the conductance scaled by the value at the critical 
concentration, which is of order ${e^2}/h$, $e$ being the electronic charge and
$h$ the 
Planck's constant, 
and $\delta_n=(n-n_c)/n_c$. Concomitantly, the data obeys scaling, 
that is, the data can be collapsed
onto two universal curves, one for the metallic samples and one for 
the insulating samples, when it is
plotted against the scaling variable
$[T_0(\delta_n)/T]^{1/x}$, where
$x$ is a  critical exponent, and the scaling parameter is 
$T_0(\delta_n)\sim
|\delta_n|^x$. The resistivity as a function of the electric field 
shows a similar scaling
behavior\cite{Krav2}, and the current-voltage  characteristic shows 
equally
striking reflection symmetry; the data on the insulating side can be 
mapped on to the metallic side by
a simple reflection. 

The immediate conclusion one can draw from these experiments is that  
the scaling and 
reflection properties are due to a
correlation length diverging at a quantum critical point\cite{Dobro} 
corresponding to a 
quantum phase transition. Moreover, the observation that the range of 
carrier
concentration over which reflection symmetry is obeyed shrinks with 
decreasing temperature is 
consistent with a zero temperature quantum critical point\cite{CHN}. 
Therefore, although the phenomenological scaling theory\cite{Dobro} 
%has indeed 
%been successful in providing a  
%picture, 
appears to be valid, the following 
%three 
key questions remain unaddressed:
%have not been addressed. 
{\em What is the origin of the quantum critical
point}? {\em What is the nature of insulating and conducting phases}?
In this paper we substantively address the origin of the quantum 
critical
point and the nature of the insulating phases, 
%but only 
and offer a conjecture regarding the conducting phase.

There have been theoretical attempts to explain the experiments in 
terms of a conventional Fermi
liquid approach\cite{Castellani} pioneered by 
Finkel'stein\cite{Finkel}.  At best, such a
perturbative renomalization group approach can provide hints 
concerning the existence of
a quantum critical point, since this point lies outside the 
perturbative regime\cite{Belitz}.
In addition, the nature of the strong coupling
phases discussed below are such that they simply cannot be accessed 
in perturbation theory.
Because the experiments do not indicate a finite superconducting 
transition temperature in
samples with
$n>n_c$, it remains to be seen if the ``metallic phase'' is actually a
superconductor in disguise\cite{Phillips,Belitz1,Zhang}. The
sensitivity of the resistivity data to  magnetic fields at the
temperatures probed in these experiments \cite{Popovic,Simonian2} 
appears inconsistent with any classical explanation.

New insight into this problem of the disordered
two dimensional electron gas can be obtained if
it is approached from the strong interaction (low
density) limit\cite{Previous};
the requisite quantum critical point 
is due to the zero temperature
continuous   melting  of a 
new state of matter, called the Wigner {\em glass}, which, as shown 
below,
is distinct from either  an Anderson insulator, or a Mott 
insulator\cite{Review}. At this 
quantum critical point scaling must hold.
In contrast,  in the absence of disorder, the insulating triangular 
Wigner crystal, which  
minimizes the Coulomb energy, quantum melts via a first order 
transition at which scaling does not
apply. Thus,  the situation is radically different with and without 
disorder. It is the principal
purpose of this paper to analyze the nature and the consequences of 
the existence  of the Wigner
glass.  

The presence of electronic spin degrees of freedom
distinguishes the Wigner glass from the similar Bragg 
glass\cite{Giam} which has been discussed in
other contexts.  The spin degrees of freedom lead to a magnetic 
Hamiltonian which  is 
highly frustrated due to the
presence of significant ring-exchange processes. Frustration, in 
turn, can lead to novel magnetic phases,
including a ``spin-liquid"\cite{Fradkin}, where a ``spin-liquid"  is 
a state  in which quantum fluctuations
are so
strong that even at $T=0$ no symmetries are broken.  As the Wigner glass melts,
the
electrons are no longer power-law ordered at the lattice sites, 
but the local crystalline character and the short distance  antiferromagnetic
spin-singlet correlations  should survive within the scale of the correlation
length, which is large
close to the quantum critical point. Most likely such a state  cannot be a Fermi
liquid due to its strong singlet correlations.
Moreover, the observed conducting state cannot be a Fermi liquid for 
it would localize if it were\cite{Dobro}. 
A simple model
of a non-Fermi liquid has a perfect metal ground state if the 
interactions
are sufficiently strong\cite{Chakravarty2}, and the interactions are 
certainly strong in the present experiments. We conjecture that the 
conducting state is a non-Fermi liquid metal with strong spin
singlet correlations.

For a Wigner glass, in the presence of disorder, it follows from the 
work of Giamarchi and Le
Doussal\cite{Giam}  that the equal-time  correlation function
\begin{equation}
C_G(R)=\overline{\langle e^{i{\bf G}\cdot{\bf u}({\bf R})}e^{-i{\bf 
G}\cdot{\bf u}(0)}\rangle}
\sim {1\over R^{\eta}},\ R\to \infty,\label{plaw}
\end{equation}
where ${\bf u}({\bf R},t)$ is the displacement  at the lattice site  
$\bf
R$, and $\eta$ is a critical exponent. The angular brackets denote 
the groundstate average and the overline the disorder average. 
Therefore, the system  exhibits quasi-long-range crystalline order 
characterized by power law Bragg
peaks at the  reciprocal lattice vectors, $\bf G$, of the triangular 
lattice,
instead 
of the $\delta$-function peaks of a crystal with true long-range 
order.  

This
quasi-long-range crystalline order  is possible only if unbound  
dislocations are not generated by  disorder.
Giamarchi and Le Doussal\cite{Giam}
have  given a self-consistent argument that this is indeed the case 
for a range of parameters, but, 
in $d=2$, this is still an open question; 
for $d=3$, this idea has received further support, 
recently\cite{Fisher,Huse}.
Pending further investigation, we shall assume the existence of the 
Bragg glass phase in $d=2$. 
Even if the self-consistent argument of Ref.~\cite{Giam} turns out to 
be incorrect,   dislocations
are expected to be  exponentially rare\cite{Huse} and so of not great 
practical importance. On
the other hand,  at very low carrier concentrations,  corresponding 
to very high disorder, 
dislocations will proliferate,  destroying this phase,  and the 
resulting
state is indistinguishable from the Mott-Anderson insulating phase. Thus, 
the
Wigner glass  should exist  only at intermediate carrier 
concentrations. 
It is also worth noting that the dissipative properties of this glass 
at the melting point should be very 
different from the melting of the  Bragg glass 
because of
the continuum of low lying fermionic  excitations.

A crucial question is whether or not  the Wigner glass is a distinct 
state of matter. 
The  findings of
Giamarchi and Le Doussal imply that it  is.
%indeed distinct. 
Specifically, it is characterized by a power-law crystalline order 
(See, Eq.~(\ref{plaw}))
and by broken replica symmetry\cite{Mezard}, 
as in a spin
glass. As a consequence, there are many low lying metastable 
states,  with barriers between them
diverging as a function of the ``separation" in the phase space. 
%Therefore, 
Such a state 
cannot be accessed by perturbative renormalization group 
methods\cite{Finkel}.

As the density is increased, the Wigner
glass will quantum melt in a continuous manner, in contrast
to the first order melting of a 
Wigner crystal; in the presence of disorder, coexistence of phases is 
not possible, and, therefore, a first order
transition is forbidden\cite{Aizenman}. However, because replica symmetry is 
restored in the melted liquid, there must 
still be a 
phase transition 
and hence a quantum critical point!

The pure Wigner crystal transition in $d=2$ is believed to occur at 
$r_s\approx 37$\cite{Ceperley}, 
where $r_s$ 
is a measure of the importance of electron-electron interactions and 
is 
defined as the average interparticle separation
in units of the effective Bohr radius $a_B=\hbar^2\epsilon/m_b e^2$; 
$\epsilon$ is the dielectric constant
and $m_b$ is the effective mass of the charge carriers. The 
situation changes dramatically in
the presence of disorder, because the elastic Wigner glass can take 
advantage of pinning and lower its
energy relative to the liquid phase.  There are  numerical results 
that indicate that the critical value of
$r_s$ can be lowered substantially, for example, to 7.5 from 37 in 
the presence of disorder\cite{Chui}. 
The
critical $r_s$ is, of course, a nonuniversal quantity and can depend 
strongly on the precise
microscopic model.

The local crystalline order of a Wigner glass allows us to estimate  
exchange energies. For 
this purpose, 
it is legitimate to treat the glass
as a triangular crystal, recognizing that the exchange constants will 
be randomly
distributed.  A general theorem shows that exchanges involving an 
even number 
of Fermions are antiferromagnetic,
while those involving an odd number of Fermions are 
ferromagnetic\cite{Thouless,Herring}.
There are estimates of these multiparticle exchanges in both solid
$^3$He and the Wigner crystal\cite{Roger}.
The version of the many dimensional WKB\cite{Schmid} formalism used 
here to calculate them is different
in detail from that of Ref.~\cite{Roger}. 
The actual
potential at short distances is, of course, softer than the Coulomb 
potential,
$e^2/\epsilon r$, due to the spread of the wavefunctions of the 
electrons perpendicular to the plane.
However, we have explicitly verified that a more realistic softer 
potential makes a negligible
difference (a little larger exchange energies) as long as the spread 
in the perpendicular direction is
less than half the lattice spacing, which is the case by a safe 
margin. This implies that our 
results are independent of the short distance microscopic details and 
depend only on the nature
of the potential on the scale of the lattice spacing, which typically 
is of order 300{\AA}.
 
The semi-classical expression for a
$p$-particle  exchange energy
$J_{p}$ is 
\begin{equation}
J_{p}=A_{p}\hbar\omega_0\left({B_{p}\over 
2\pi\hbar}\right)^{1/2}e^{-B_{p}/\hbar}.
\label{exchange}
\end{equation}
Here $B_{p}$ is the value of the Euclidean action along the minimal 
action path that  exchanges $p$
electrons on a triangular lattice, and $\omega_0$ is the attempt 
frequency, 
which  can be estimated from the  
phonon
spectrum of the Wigner lattice\cite{Phonon}. 
We ascribe the zero point energy 
of the phonons to an
effective  oscillator, $\hbar\omega_0=1.63/r_s^{3/2}$ in units of the 
Rydberg constant, 
which is
$e^2/2\epsilon a_B$.  The prefactor $A$, in  one dimensional WKB 
problems,
can be proven to be greater than unity\cite{Schwinger} and is often 
quite a bit larger 
than unity; this should  hold more generally. We set $A_{p}=1$, which 
is an underestimate. Equation 
(\ref{exchange}) can be trusted as long as
$B_{p}/\hbar$ is much larger than unity.   

We have calculated $B_{p}/\hbar$ for a large number of exchange 
configurations (upto $p=18$) and have also made  asymptotic
estimates  for $p\gg 1$. Here, we report results for only the most 
compact 2, 3, 4, 5, and 
6-particle exchanges for the Coulomb potential, of which 5 and 
6-particle exchanges were not previously
calculated. This required minimization over typically 500-1000 
variables to obtain satisfactory
convergence. If we define $B_p/\hbar=b_p r_s^{1/2}$, then 
$b_2=1.66 $, $b_3=1.52$, $b_4=1.67$,
$b_5=1.91$, and $b_6=1.77$. The values of $b_p$ for $p=2, 3$, and $4$ 
are typically in the range 10-15\%
lower than the previously known estimates\cite{Roger}; this is 
because full minimization was not
previously carried out. 

If we retain for simplicity only exchanges up to 4 particles, it is 
easy to show in terms of the
spin-1/2 operators,
$\bf S$, that for the perfect triangular lattice, the spin 
Hamiltonian is 
\begin{eqnarray}
H={\rm cst.}&+&J_{NN}\sum_{\langle ij \rangle}^{NN}{\bf S}_i\cdot{\bf 
S}_j\nonumber \\
&+&J_{NNN}\sum_{\langle ij\rangle}^{NNN}{\bf
S}_i\cdot{\bf S}_j+4J_4\sum_{\langle ijkl \rangle}^{Rhombus} G_{ijkl},
\end{eqnarray}
where $G_{ijkl}=({\bf S}_i\cdot{\bf S}_j)({\bf S}_k\cdot{\bf 
S}_l)+({\bf S}_i\cdot{\bf S}_l)({\bf S}_j
\cdot{\bf S}_k)-({\bf S}_i\cdot{\bf S}_k)({\bf S}_j\cdot{\bf S}_l)$. 
The first sum is 
%a sum 
over all
distinct nearest neighbors, 
where $J_{NN}=4(J_2+{5\over 4}J_4-J_3)$. 
The second sum is over
all distinct next nearest neighbors,
where $J_{NNN}=J_4$. The third sum is 
%a sum
over all distinct rhombuses. 
$J_{NN}$ is antiferromagnetic for $r_s\lesssim 39$, and ferromagnetic 
for larger $r_s$. Its magnitude, using $\epsilon=7.7$ and $m_b=0.2$ 
(appropriate for Si-MOSFET) is 5.2 K, 1.6 K, 0.4 K, 0.06 K for 
$r_s=5, 7, 10$, and 15, respectively. Note, however, 
that while $J_{NNN}/J_{NN}\approx 0.4$ at $r_s=5$, the ratio 
approaches 1 at
about $r_s=29$. This is a highly frustrated antiferromagnetic spin 
Hamiltonian. 
Numerical evidence\cite{Lhuillier} suggests that the ground state of 
this Hamiltonian is ``quantum
disordered" for $r_s\lesssim 39$, where ``quantum disordered" means 
that spin-rotational symmetry is
not broken, even at $T=0$. Because the actual Wigner glass  is not 
translationally invariant,
any quantum disordered state is a spin liquid; various dimerized 
states, which break translational
symmetry, are not distinct states of matter. However, 
because of disorder, the spin liquid only exhibits a spin pseudo-gap state, 
rather than the spin gap state of the pure system\cite{Lhuillier}.
Other magnetic 
structures may develop for
$r_s\gtrsim 39$, but they are likely to be predominantly 
ferromagnetic.  

In the current experiments
cited earlier, the critical values of $r_s$ at which the metal-insulator 
transition takes place are in
the range of 5-15. (The character of this transition will be very 
different if it takes place for larger
values of $r_s$ for which exchange is  of insignificant importance at 
experimentally realistic
temperatures.) Because the exchange energies at
$r_s\sim 10$ are about 0.4 K, the spin-singlets can be broken with 
modest  magnetic 
fields of order 1T, and
the ground state should be very sensitive to such fields. Once the spin-singlets
are broken, the elctrons can be described as a Fermi liquid, which should
localize in 
$d=2$ due to disorder. 
Indeed, it has been observed  that the metallic behavior in Si MOSFETs can be 
suppressed by  such fields\cite{Simonian2}. In the absence of the magnetic
field,
the continuous nature
of the  melting of the Wigner glass allows the spin correlations on the scale
of 
the correlation length (large close to the melting transition) to be similar 
to those in the Wigner glass. In particular, the spin pseudogap character 
should survive the melting process, and the  drop of
the resistivity as a function of temperature seen in  experiments is due to the
reduction of the 
scattering rate of the carriers as the temperature is
lowered due to the opening of the spin pseudogap.

\begin{figure}[htb]
\centering
\epsfxsize=\linewidth
\epsffile{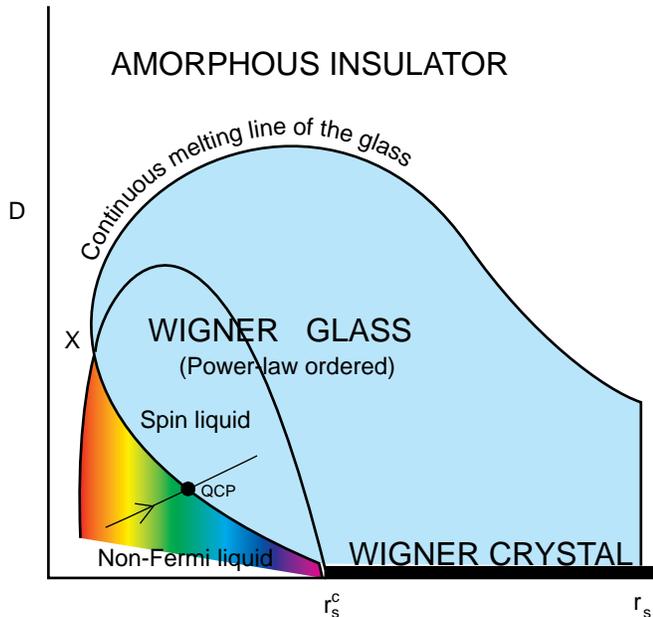}
\caption{The $T=0$ phase diagram as a function of disorder $D$ and 
$r_s$. The point $X$ is a tetracritical point. The 
arrowed curve represents the postulated experimental
path; QCP is the corresponding quantum critical point.
The Amorphous Insulator crosses over from an Anderson-Mott
description at low $r_s$ to Efros-Shklovskii behaviour
at large $r_s$.}
\label{Fig}
\end{figure}

The qualitative global phase diagram at $T=0$ is sketched in 
Fig.~\ref{Fig} with $D$ as the measure of the strength
of the quenched disorder. 
 All the transition 
lines are continuous quantum phase transitions for which scaling  
holds. The point $X$ 
is a tetracritical
point, while $r_s^c$ is the first order transition of the pure 
two-dimensional electron gas to a 
Wigner crystal
at which scaling does not hold. In a Coulomb system, the density is
determined by the charge-neutrality condition. Hence, the density is 
continuous;
only the order parameter is discontinuous.
The spin-liquid is a phase within the Wigner glass. 
The Fermi liquid behavior of
the $D=0$ line is unstable with respect to arbitrarily weak disorder. 
It either
flows towards the Mott-Anderson insulating state or to a non-Fermi liquid 
state. 
It is thus possible that
the Mott-Anderson insulator to non-Fermi liquid phase boundary terminates 
in the $D\to 0$ limit at the origin ($D=r_s=0$),
or it may terninate at a point of finite interaction strength ($D=0$, 
$r_s=r_s^A$). Also, note the reentrance of the
Mott-Anderson insulator as a function of $r_s$.

The Wigner glass phase 
is stable to small thermal perturbations 
and must
exhibit a finite temperature phase transition with a transition 
temperature that vanishes at the
quantum critical point. However, neither the Mott- Anderson insulator, nor the
non-Fermi
liquid are distinct phases of matter at finite temperatures. 
 Because of the continuous Heisenberg symmetry of the spin
Hamiltonian, there cannot be any magnetic long range order
in $d=2$ at any finite
temperature.   
The transitions
shown in the figure are true thermodynamic phase transitions at 
$T=0$, and are, in principle,
detectable as non-analyticities in thermodynamic functions, such as 
the chemical potential. A prediction is that the compressibility, $\kappa$, of
the electrons will vanish as $\kappa\sim (n-n_c)^{\nu}$, which should be
experimentally
observable in capacitive measurements. (As defined earlier, $\nu$ is the correlation length 
exponent that has been determined experimentally to be 1.5 from the temperature and the electric field
scalings\cite{Krav2}.) The argument is quite simple. From  macroscopic electrodynamics, the dielectric
function, $\epsilon(q,\omega=0)=1+ {2\pi e^2\over q}\kappa(q)$. In the insulating Wigner glass state,
the $q=0$ dielectric function
is finite and different from that 
of the vacuum. This implies that
$\kappa(q)\sim q$ as $q\to 0$.
Since $\kappa(q,\xi)$ should be continuous across $q\xi\sim 1$,
it follows from the scaling hypothesis that $\kappa(q,\xi)\sim {1\over \xi}$
in the critical regime, $q\xi >1$.
The critical behaviour of the compressibility should be
observable in macroscopic measurements 
on the conducting side of the transition. 

The
magnetic transitions will be reflected in the magnetic response of 
the electron
spins as  $T\to 0$.  In particular, a signature of the the 
spin-liquid
to ferromagnetic phase transition can be detected in the uniform 
susceptibility, which will diverge as the transition is approached from
the spin-liquid side.  A similar signature may be observable
in the uniform susceptibility at the non-Fermi-liquid to Mott-Anderson
insulator transition.  Because
of the power-law crystalline order of the Wigner glass, its presence 
can also be detected in surface
acoustic wave and narrow band noise measurements.

By applying electric field, a  Wigner
glass can be made to slide\cite{Millis}.  In general, there should 
not be any associated threshold 
field; although the equilibrium Wigner glass may not contain 
dislocations, these will be generated in
the presence of an external electric field. The mechanism which 
destroys any possible threshold field is similar
to that discussed by Coppersmith\cite{Coppersmith}. The 
current-voltage characteristic in  the glass
phase is $J\sim E \exp(-(E_c/E)^{\mu})$, where $E$ is the electric 
field, $J$ 
is the current density,  $E_c$ is the {\em typical} pinning field, 
and $\mu$ is an exponent.
This reflects the presence of infinitely many low  lying metastable 
states.  As the glass melts,
the current voltage characteristic should become linear at 
asymptotically low currents. Close to the
transition, the scaling picture suggests that the conductivity 
can be written as 
$\sigma(\omega,E,T)=\sigma\left({\delta_n\over\omega^{1/\nu z}}, 
{\delta_n\over E^{1/\nu (z+1)}}, {T\over \omega}\right)$, where $\nu 
z$ is precisely the exponent $x$ 
defined earlier. 
The glass correlation length is given by 
$\xi_g\sim \delta_n^{-\nu}$, and $z$ is the dynamic critical exponent 
defined by $\xi_{\tau}\sim
\xi_g^z$, where $\xi_{\tau}$ is the correlation time.

We emphasize, once again, that
for  purer systems, the metal-insulator
transition must approach the first order 
melting  of a Wigner crystal at which  scaling cannot hold, while
in  more disordered samples  scaling will hold. 
The scaling
theory\cite{Dobro} of the metal insulator transition involves  a 
transition between
an insulator and a perfect metal.  This description is valid
in the critical regime, even if the actual asymptotic behavior of
the ``metallic'' phase is  different from that of a perfect metal;
outside the critical regime, the behavior of the ``metallic'' phase 
can
be affected by couplings that are irrelevant at the critical
fixed point (and so do not affect the critical phenomena), but are 
relevant
at the putative stable fixed point.  Thus, it is perfectly consistent 
with
the scaling theory for the resistivity outside the
critical regime to saturate at low temperatures, as has been
observed in some experiments, or even to turn around and ultimately 
diverge
at very low temperatures.

We  hope that many
aspects of our proposed phase diagram can be explored, both
theoretically and experimentally, in the near future.  It is
also worthwhile to consider the extension of the present ideas to
three dimensional systems, where the existence of a Wigner glass
phase is on even firmer theoretical footing.

{\it Note added}: We thank V. M. Pudalov for drawing our attention to
earlier evidence of a Wigner crystal transition in Si MOSFETs described
in V. M. Pudalov, {\it et al.}, {\sl Phys. Rev. Lett.} {\bf 70}, 1866 
(1993).
However, from the perspective of the present paper, this transition
is better described as a Wigner glass transition with no threshold 
field.

After this paper was submitted for publication, further evidence of 
a Wigner glass  transition was discovered (J. Yoon {\sl et al.},
cond-mat/9807235). 
The authors describe
this as a Wigner crystal transition from the proximity of the transition to
$r_s^c=37$ obtained from fixed-node quantum Monte Carlo approximation.
This interpretation is incorrect as the disorder 
is finite, which by all theoretical accounts must destroy the Wigner
crystal. We believe that they have observed a transition to the Wigner
glass state with a power-law order as described here, close, however, to the 
first order melting of the pure crystal. It is  therefore not surprising
that scaling is approximately violated; asymptotically close to the transition
scaling 
must of course hold as the system is strictly speaking not pure.

H.-W. Jiang in private communications has reported to us that the
compressibility
in his hole doped GaAs samples indeed vanishes as a power-law 
at $r_s^c\sim 8$ as described above.

The evidence of the transition to an insulating state at smaller values of
$r_s$ as
shown in Figure 1 has been found (A. R. Hamilton {\sl et al.},
cond-mat/9808108).
The character of this second transition is very different from that of the
first transition.
This implies that there must be at least two distinct insulating states, as
there cannot
be two different critical transitions to the same insulating state
from one conducting state. We suggest that this lends
further support to our phase diagram.


\begin{references}
\bibitem{Previous}That the low density phase can be a  somewhat 
disordered
Wigner crystal, has been mentioned in the past on a number of 
occasions (more recently in
Refs.~\cite{Phillips,Zhang}). However, none of the striking aspects 
of the Wigner glass discussed here
were addressed.
\end{references}
\end{document}